\documentclass{mn2e}
\usepackage{epsfig}
\def\msun{{\rm M_{\odot}}}

\title[Variability in black hole accretion discs] 
{Variability in black hole accretion discs} 
\author[A.R.~King, J.E.~Pringle, R.G.~West and
M. Livio]{A.R.~King$^1$, J.E.~Pringle$^{2,3}$, R.G.~West$^1$ and M. Livio$^3$\\
$^1$Department of Physics and Astronomy, University of Leicester,
Leicester, LE1 7RH, UK\\
$^2$Institute of Astronomy, Madingley Rd, Cambridge, CB3 0HA, UK\\
$^3$Space Telescope Science Institute, 3700 San Martin
Drive, Baltimore, MD 21218, USA}
\date{Submitted: July 2003}

\pagerange{\pageref{firstpage}--\pageref{lastpage}} \pubyear{2003}

\begin{document}
\def\lta{\mathrel{\spose{\lower 3pt\hbox{$\mathchar"218$}}
     \raise 2.0pt\hbox{$\mathchar"13C$}}}
\def\gta{\mathrel{\spose{\lower 3pt\hbox{$\mathchar"218$}}
     \raise 2.0pt\hbox{$\mathchar"13E$}}}
\def\Msun{{\rm M}_\odot}
\def\msun{{\rm M}_\odot}
\def\Rsun{{\rm R}_\odot}
\def\Lsun{{\rm L}_\odot}
\def\19{GRS~1915+105}
\label{firstpage}
\maketitle

\begin{abstract}

Observations of accreting systems often show significant variability
(10--20 percent of accretion luminosity) on timescales much longer
than expected for the disc regions releasing most of the luminosity.
We propose an explicit physical model for disc variability, consistent
with Lyubarskii's (1997) general scheme for solving this problem. We
suggest that local dynamo processes can affect the evolution of an
accretion disc by driving angular momentum loss in the form of an
outflow (a wind or jet). We model the dynamo as a small--scale
stochastic phenomenon, operating on roughly the local dynamical
timescale. We argue that large--scale outflow can only occur when the
small--scale random processes in neighbouring disc annuli give rise by
chance to a coherent large--scale magnetic field. This occurs on much
longer timescales, and causes a bright large--amplitude flare as a
wide range of disc radii evolve in a coherent fashion. Most of the
time, dynamo action instead produces small--amplitude flickering. We
reproduce power spectra similar to those observed, including a $1/f$
power spectrum below a break frequency given by the magnetic alignment
timescale at the inner disc edge. However the relation between the
black hole mass and the value of the break frequency is less
straightforward than often assumed in the literature. The effect of an
outer disc edge is to flatten the spectrum below the magnetic
alignment frequency there. We also find a correlation between the
variability amplitude and luminosity, similar to that found in some
AGN.

\end{abstract}

\begin{keywords}
black hole physics -- X--rays: binaries, galaxies: jets
\end{keywords}

\section{Introduction}

Both Galactic and extragalactic accretion--powered X--ray sources
display significant aperiodic variability on a wide range of
timescales. Similarities in behaviour between the black hole candidate
X--ray binaries (BHCs) and the nuclei of Seyfert galaxies (that
presumably also harbour central black holes) have been used to suggest
that the physical processes occuring in accretion flows around the
black holes in these objects are essentially the same, but scaled in
some appropriate manner. There are indeed suggestions that such scaling
relationships might be used as an estimate for black hole masses
(e.g.\ Vaughan, Fabian \& Nandra, 2002; Markowitz et~al.\ 2003). Although
X--ray binaries containing neutron stars as the accreting compact
objects also exhibit interesting aperiodic, and quasi-periodic, behaviour
(see for example the review by van der Klis 1989), we restrict our
attention in this paper to black hole accretion discs, partly in the
hope that understanding the physics of such accretion flows might be
somewhat less complicated, but also because of the interest in scaling
between stellar and supermassive objects.

We stress, however, that the processes we discuss in this paper are
fully relevant not only to binary X--ray sources and AGN, but to all
objects containing accretion discs, such as cataclysmic variables and
protostellar objects.

Two main diagnostics are relevant for the variability of these
sources. These are the shape of the power spectrum and the amplitude
of the variability. Of course, the latter can be deduced from the
normalisation of the former (e.g.\ Miyamoto et~al.\ 1992) and so strictly
speaking, these are just two aspects of the same diagnostic.

The best observed black-hole candidate X--ray binary (BHC) is Cygnus
X--1. In the low/hard state the normalised rms variability,
$\bar{\sigma}$, is of order a few tens of per cent and the power
spectrum can be approximated as a broken power law of the form $p(f)
\propto f^{-\alpha}$ where $\alpha \simeq 0$ for $10^{-3}$~Hz $< f <
0.1$~Hz, $\alpha \simeq 1$ for $0.1$~Hz $< f < f_{\rm br} \simeq
3$~Hz, and $\alpha \simeq 2$ for $ f_{\rm br} < f < 10^3$~Hz (Nowak 
et~al.\ 1999; Revnivtsev, Gilfanov \& Churazov 2000; Churazov, Gilfanov
\& Revnivtsev 2001). This implies that the flickering variability
occurs predominantly on timescales in the range 0.3~s -- 10~s. In the
high/soft state the normalised rms variability, $\bar{\sigma}$, is of
the order of a few per cent, and the spectrum is a similar, broken
power law, except that now $f_{\rm br} \simeq 10$~Hz, and $\alpha
\simeq 1$ for $f < f_{\rm br}$. In this state, therefore, the
flickering occurs predominantly on timescales $\ga 10$~s. Other BHCs
show similar behaviour. The power spectra of GX339-4 and GS2033+338 in
the low/hard state are given by Miyamoto et~al.\ (1992) who note their
strong similarity to Cyg X--1. The power spectrum of LMC X--1 is shown
by Nowak et~al.\ (2001) to be similar to Cyg X--1 in the high/soft
state, and to have a normalised rms variability of $\bar{\sigma}
\simeq 7$ per cent.

X--ray power spectra of Seyfert 1 galaxies also typically display a
broken power law, with the break occurring at frequency $f_{\rm br}
\simeq 10^{-6}$~Hz, corresponding to a timescale of the order of a few
days (Markowitz et al. 2003). For $f < f_{\rm br}$ the slope of the
spectrum has typically $\alpha \simeq 1$ and for $f > f_{\rm br}$ the
slope of the spectrum has typically $\alpha \simeq 2$ (see also
Vaughan \& Fabian 2003; Vaughan et al.\ 2003). The normalised rms
variability amplitudes range typically up to $\bar{\sigma} \simeq
10$ per cent (Markowitz et~al.\ 2003).

Uttley and McHardy (2001) have drawn attention to a linear correlation
between the actual rms variability, $\sigma$, and the flux, $F$, in
both BHCs and Seyfert~1 galaxies, implying that the normalised rms
variability $\bar{\sigma}=\sigma/F$ is approximately independent of 
source brightness, at least over a small range in flux. From the data 
for the BHCs Cyg X--1 and SAX J1808.4-3658, given in the paper by Uttley 
\& McHardy (2001), we find that the relationship is approximately of the 
form $\bar{\sigma} \propto F^q$ where $q \simeq 0.3$.

The earliest models for the variability were in terms of shot noise
(Terrell 1972); the light curves were reconstructed from a series of
independent overlapping shots. Such a picture, or variations thereof
(e.g.\ Lehto 1989), can be made to model the data reasonably well, in
particular the shape of the power spectrum. However, this scenario
still lacks a physical picture for the variability. Since then,
various ideas have been put forward, including accretion disc
turbulence (Nowak \& Wagoner, 1995), magnetic flares in the disc
corona (e.g.\ Poutanen \& Fabian, 1999), and MHD instabilities in the
plunging region of the inner disc (Hawley \& Krolik, 2001). These
models all have a common problem in accounting simultaneously for the
range of timescales and the normalised rms variability amplitudes. The
latter are typically around 10 per cent, and imply emission from the
inner few disc annuli, where most of the accretion energy is released
(see for example the discussion by Bruch, 1992; Fritz \& Bruch,
1998). But these models also imply that characteristic variability takes
place on local dynamical timescales. At the inner radii of accretion
discs around $\sim 10 M_{\sun}$ black holes, these relevant timescales
are far too short, typically of the order of milliseconds.

An insightful paper by Lyubarskii (1997) offers a way out of these
difficulties. Lyubarskii considers small amplitude local fluctuations
in the accretion rate at each radius, caused by small amplitude
variations in the viscosity, and then considers the effect of these
fluctuations on the accretion rate {\it at the inner disc edge}. A
linear calculation shows that if the characteristic timescale of the
viscosity variations is everywhere comparable to the viscous (inflow)
timescale, and if the amplitude of the variations is independent of
radius, then the power spectrum of luminosity fluctuations has the
form $p(f) \propto 1/f$. If the amplitude of the variations increases
with radius, the slope of the power spectrum of the luminosity
variations is steeper than $-1$. Lyubarskii (1997) notes that he has
no physical model for the cause of such fluctuations. In particular,
although the obvious candidate cause is the magnetic dynamo, the
characteristic timescales for the dynamo are much shorter than the
local viscous timescale.

In this paper we use the idea (Livio, Pringle \& King 2003) that
although the local dynamo timescale at each point in the disc is
indeed short, the timescale for the dynamo processes in a sufficient
number of independent neighbouring disc annuli to produce a poloidal
magnetic field coherent enough to affect the accretion rate, by
generating a disc wind or jet, can be quite long. That a disc wind/jet
can be a significant driver of accretion in such objects is in line
with the recent ideas about `jet--dominated' states in black hole
candidate binaries proposed by Fender, Gallo \& Jonker (2003). We
propose a specific, but highly simplified, model of an accretion disc.
A magnetic dynamo process generates the viscosity, but also, from time
to time, produces a sufficiently well--ordered poloidal
field. Therefore, the angular momentum is transferred outwards by two
separate processes --- the usual magnetic viscosity and a disc wind or
jet. Numerical simulation of all the physical processes involved in
such an intricate system is, of course, unrealistic at the current
time, and we make no attempt to describe the full MHD problem in its
three--dimensional complexity (e.g.\ Hujeirat, Camenzind \& Livio
2002).  Instead, we make a number of simplifying assumptions about the
physical processes involved, and use these to construct a numerical
description of the time-dependence of such an accretion disc. We show
below that this straightforward approach, based on simple physical
ideas, can describe the flickering behaviour observed in accreting
black hole systems.

\section{The Equations for Disc Evolution}

In this Section we derive and explain the equations we use to describe
the time evolution of the disc. We simplify the problem by adopting a 
one--dimensional formulation and
using current ideas and simulations to introduce some of the basic
physics. We measure disc radii in units of the inner radius $R_{\rm
in}$, and assume an outer radius $R_{\rm out}$ large enough (typically
$20,000R_{\rm in}$) that the outer disc can act as a reservoir, allowing
the inner disc to approach a quasi-steady state. We avoid undue
complexity in these preliminary computations by ignoring the details of
internal structure and thermal effects within the disc. Instead, we
assume a fixed disc opening angle, with thickness $H \propto$ disc
radius $R$.
We typically take $H/R = 0.08$ for numerical convenience: a smaller
and perhaps more realistic value would require longer run times,
without changing the character of our results.
We use a fixed grid with
logarithmic spacing. We choose the zone width $D\!R \simeq H$ and thus
$N \simeq 130$ zones. This is not strictly necessary, but gives
sufficient accuracy for our current purposes --- we note that the main
errors in our treatment are likely to come more from our assumptions
than from numerical inaccuracy. Indeed, using a formal numerical radial
resolution which is finer than the disc thickness, is not likely to
have physical meaning. In addition, the convenience of the choice $D\!R
\simeq H$ becomes apparent in Section~\ref{dynamo}. The disc material
is assumed to have Keplerian angular velocity $\Omega =
(GM/R^3)^{1/2}$ about a central point mass $M$, and we use units with
$GM = 1$. 
In the present paper we consider a disc extending inwards close to the
last stable orbit around the black hole, as is generally thought to
occur in the high/soft X--ray state.

\subsection{The surface density}

We assume that the surface density $\Sigma$ evolves because of viscous
angular momentum transfer within the disc and because of angular
momentum loss in the magnetic wind or jet. Note that we use the terms
jet and wind simply to mean a mechanism by which the disc can lose
energy and angular momentum, without being more specific. Thus we
write (Pringle, 1981; Livio \& Pringle 1992)
\begin{equation}
\frac{\partial \Sigma}{\partial t} = \frac{3}{R}
\frac{\partial}{\partial R} \left[ R^{1/2} \frac{\partial}{\partial R}
(\nu \Sigma R^{1/2}) \right] - \frac{1}{R} \frac{\partial}{\partial R}
\left[ R \Sigma U_R \right],
\label{eqsigma}
\end{equation}
where $U_R < 0$ is the radial velocity induced by angular momentum
loss in the jet/wind (see Section~\ref{jettorque}). Note that we
assume that although the jet/wind removes angular momentum from the
disc, it does not remove a significant amount of material. It is clear
from global energy considerations that the wind cannot remove all of
the disc material (put simply, all of the accretion energy cannot go
into driving all of the accreting material back to infinity). Hence,
at the level of accuracy to which we are working, we neglect
mass--loss to the wind in the surface density evolution.

The viscosity $\nu$ is assumed to result from magnetic torques
generated by dynamo activity within the disc. We use here the standard
Shakura-Sunyaev (1973) parametrisation of that viscosity using the
dimensionless parameter $\alpha$ in the form
\begin{equation}
\nu = \alpha c_s H,
\end{equation}
where $c_s$ is the (appropriately averaged) sound speed in the
disc. Using hydrostatic balance we have approximately
\begin{equation}
c_s = H \Omega,
\end{equation}
and hence
\begin{equation}
\nu = \alpha H^2 \Omega.
\end{equation}
Thus if there is no loss of angular momentum to the wind, (i.e. $U_R
=0$), then the evolution equation is a linear diffusion equation for
the surface density $\Sigma$. We also note that by using this simple
prescription and ignoring the thermal structure of the disc, we have
for the time being put aside discussion of the role that hysteresis
loops in the $(\nu, \Sigma)$-plane might play. These are the well-known 
S--curves which are thought to give rise to the outbursts of cataclysmic
variables and X--ray transients, and might well play a role in such
systems (e.g.\ Belloni et~al.\ 1997; Watarai \& Mineshige, 2003). In
this paper we consider disc instabilities caused by the interaction of
viscous and wind torques alone.

We take a zero torque condition $\Sigma = 0$ at the inner boundary
$R=1$, allowing accretion to occur freely through the inner
boundary. At the outer boundary we impose a zero radial velocity
condition. We use a simple first order explicit numerical scheme,
using upwind derivatives for the advective $U_R$ term. The scheme is
written so that, except at the inner boundary, disc mass is conserved
to machine accuracy.

\subsection{The magnetic field}

At each point in the disc we postulate a poloidal (vertical) field
component $B_z$ generated by the disc dynamo producing the viscosity
$\nu$. We assume that this field drives the magnetic wind torques
causing the inflow velocity $U_R$. We assume that the field is
advected by $U_R$, and that it is also able to diffuse through the
disc because of the magnetic diffusivity $\eta^\ast$ resulting from
the dynamo process. We discuss each of these assumptions below when
describing their implementation in our disc evolution scheme. They
imply an evolution equation for $B_z$ in the form
\begin{equation}
\frac{\partial B_z}{\partial t} = - \frac{1}{R}
\frac{\partial}{\partial R} \left( R B_z U_R \right) +\frac{1}{R}
\frac{\partial}{\partial R} \left( R \eta^\ast \frac{\partial
B_z}{\partial R} \right) + {\cal S}_B.
\end{equation}

We should note here that our scheme neglects the advection of magnetic
field by the velocity induced by the viscosity $\nu$. The dynamo
process generating $B_z$ in random fashion (Section~\ref{dynamo}) is
an integral part of the process generating viscous torques, and hence
viscously driven accretion. This allows diffusion of $B_z$
radially through the disc (Section~\ref{magdiff}). But, for unit
Prandtl number, the timescales on which $B_z$ varies locally due to
the dynamo ($k_d \Omega^{-1}$, where physically we expect $k_d \sim
10$, Section~\ref{dynamo}) and due to diffusion between neighbouring
disc annuli ($\sim H^2/\eta \sim \alpha^{-1} \Omega^{-1}$, 
Section~\ref{magdiff}) are typically much shorter than the radial 
viscous diffusion timescale ($\tau_{\rm visc}\sim R^2/\nu$).

At the outer disc boundary we take $B_z =0$, allowing magnetic flux to
diffuse outwards, but preventing inward flux advection. At the inner
disc radius we assume zero flux loss through either diffusion or
advection. The scheme is written so that poloidal flux is conserved to 
machine accuracy through the disc, accounting for diffusion through the 
outer boundary and the dynamo source term.

\subsubsection{The source ${\cal S}_B$ of $B_z$}
\label{dynamo}

We assume that the source of the local vertical field component $B_z$
is the disc dynamo process driving the viscosity.  Specifically we
assume that, in this respect, each annulus of the disc of width $D\!R
\approx H$ acts independently of all other annuli. We further assume
that each annulus generates large enough fields $B_{\rm disc}$ internal to
the disc to produce the pseudo--viscous magnetic torques contained in
the $\alpha$--parameter. That is, we assume (Shakura \& Sunyaev 1973)
\begin{equation}
\alpha = \frac{B_{\rm disc}^2}{(\Sigma/2H) \, c_s^2},
\end {equation}
and also that the dynamo process gives rise to a local
vertical field $B_z$ of magnitude $\la B_{\rm disc}$. We assume, that for
each annulus, $B_z$ can be modeled in terms of a stationary series
resulting from a stochastic process. We model this process in terms of
the Markoff scheme (e.g.\ Kendall 1976). This is the simplest linear
autoregressive scheme other than a purely random series. In this
scheme the $n$--th member of the series $u_n$ is given in terms of the
previous by the relation
\begin{equation}
u_n = - \alpha_1 u_{n-1} + \epsilon_n,
\end{equation}
where $\alpha_1$ is a parameter, which for stability has modulus less
than unity, and $\epsilon_n$ is a random variable with zero mean.

This scheme produces a Markoff series with oscillations
of a more or less regular kind. The mean number of series steps
between oscillation peaks is
$2 \pi/\cos^{-1} [-(1 + \alpha_1)/2 ]$, and the amplitude of
the oscillations is such that

\begin{equation}
{\rm var} \, u = \frac{{\rm var} \, \epsilon}{1 - \alpha_1^2}.
\end{equation}

In our simulations we choose $\alpha_1 = -0.5$, so we expect 3.45
series steps between peaks. We envisage the dynamo process to have
some local canonical timescale $\tau_d(R)$.  We assume that $\tau_d(R)
= k_d \Omega^{-1}$ with $k_d \sim 10$ (e.g. Tout \& Pringle 1992; 
Stone et~al.\ 1996). We therefore apply the Markoff process
through the source term ${\cal S}$ with the appropriate frequency at
each radius so as to give rise to that mean timescale.

We vary the amplitude of the process by adjusting the variance of the
term $\epsilon_n$. To do this we use a dimensionless parameter
$B_s$. Then $\epsilon$ is chosen to be a random variable in the range
$[ B_{\rm max},-B_{\rm max} ]$. The value of $B_{\rm max}$ is
proportional to $B_s$ and chosen so that if $B_s =1$ the resulting
$B_z$ would, on average, produce a magnetic torque
(see~\ref{jettorque}) driving an inflow velocity similar to that
generated by the viscosity, i.e.\  $\nu /R$. Thus we take

\begin{equation}
B^2_{\rm max} = \frac{3}{2} B^2_s \, \pi \alpha \Sigma \, \frac{H}{R}
\, \frac{GM}{R^2}.
\end{equation}

\subsubsection{The jet torque and $U_R$}
\label{jettorque}

We assume that the poloidal magnetic field generates a jet/wind
removing angular momentum and energy, but negligible mass, from the
disc, and so produces inflow velocity $U_R$. One way to proceed
(cf.\ Lovelace, Romanova \& Newman 1994; Lovelace, Newman \&
Romanova 1997) would be to assume that the inflow velocity in the
disc is simply proportional to $B_z^2$. However, since the source of
the poloidal field is the dynamo process which has a radial scale in
the disc of order $\sim H \ll R$ we feel that this is not physically
justifiable. Instead, we follow the ideas of Tout \& Pringle
(1996; see also Livio 1996). They suggest that an inverse cascade driven 
by disc dynamics and reconnection can transform these small--scale 
poloidal fields into a global poloidal field smaller in magnitude but 
larger in scale. The component of the poloidal field with a scale of 
order $\sim R$ could then produce a magnetically dominated outflow.

To model these ideas in the current simulation we define two
quantities 
\begin{equation}
\langle B_z \rangle \equiv \frac{\int_{R - \Delta/2}^{R + \Delta/2}
B_z R dR}{R \Delta},
\end{equation}
and
\begin{equation}
\langle B_z^2 \rangle \equiv \frac{\int_{R - \Delta/2}^{R + \Delta/2}
B_z^2 R dR}{ R \Delta},
\end{equation}
where typically we take the range $\Delta \simeq R$. Then the quantity
\begin{equation}
\label{Qdef}
{\cal Q}(R) = \langle B_z \rangle^2 / \langle B_z^2 \rangle
\end{equation} 
is a dimensionless measure of the local coherence of the
dynamo-generated field, and is such that $0 \leq {\cal Q} \leq 1$.

We could now simply allow the local magnetic torque to depend on
$\langle B_z \rangle$ rather than $B_z$ (cf.\ also Livio \& Pringle,
1992) and assume that it gives rise to an inflow velocity, say $V_R$,
given by

\begin{equation}
V_R = - \frac{\langle B_z \rangle^2 R^{3/2}}{ (GM)^{1/2} \pi \Sigma}.
\end{equation}

However, we must also consider the ability of a poloidal field to
launch a wind or jet. Although a sufficiently large--scale poloidal
field is probably able to give rise to a wind of some sort, Blandford
\& Payne (1982) point out that the wind is considerably enhanced if
the poloidal field lines make a large angle with the vertical, i.e.\ 
if $B_R/B_z$ is large enough (see also Ogilvie \& Livio 2001).
Such a field configuration results if the inflow velocity is
comparable to the speed, $\sim \eta^\ast /R$, with which vertical
field can diffuse outwards through the disc. Here $\eta^\ast$ is the
effective magnetic diffusivity in the disc, defined below in
Section~\ref{magdiff}. This effect is discussed
by Lubow, Papaloizou \& Pringle (1994b) and by Lovelace, Romanova \&
Newman (1994) who conclude that it can drive strongly time--dependent
disc behaviour.  These ideas are confirmed by the `accretion-ejection'
events seen in the numerical calculations of Casse \& Keppens
(2002). Lubow et~al.\ (1994b) suggest that the strength of the wind is a
sensitive function of the quantity $X = R V_R/\eta^\ast$. Lubow et~al.\
(1994b) argue that for a locally isothermal disc the loss from the
disc is $\propto \exp (-1/X^2)$, and that for large $X$ the loss
timescale is the local dynamical one. In the simulations reported here
we do not use the non--analytic function $\exp(-1/X^2)$ (sometimes
called the April Fool Function, because, although all its derivatives
exist at the origin, it is not equal to its null Taylor series about
that point). Instead, for numerical reasons we approximate the
sensitivity to $X$ by assuming that the radial velocity driven by the
jet/wind obeys

\begin{equation}
U_R = V_R \left( 1 + \{f_{\rm LPP} -1 \} \frac{X^2}{X^2 + 4} \right).
\label{x}
\end{equation}
Here the parameter $f_{\rm LPP}$, assumed $\geq 1$,
represents the maximum enhancement to the torque that this effect can
achieve. This formula has the behaviour that as
$X \rightarrow \infty$, $U_R \rightarrow f_{\rm LPP} V_R$, and $U_R$
tends to $V_R$ at small $X$. Note that setting $f_{\rm LPP} = 1$
removes this effect.

\subsubsection{The magnetic diffusivity $\eta^\ast$}
\label{magdiff}

We assume that the disc dynamo producing the effective viscosity $\nu$
also gives rise to an effective magnetic diffusivity $\eta$. It seems
reasonable to assume that the Prandtl number is of order unity, and
thus that $\eta = \nu$. However, if the poloidal field is aligned over
a scale of order the radius $R$ then it diffuses through the disc at a
rate enhanced roughly by a factor of $R/H$ (van Ballegooijen 1989;
Lubow, Papaloizou \& Pringle 1994a). Of course, the actual rate at
which the poloidal field diffuses through the disc depends on the
details of the configuration of the field which we do not attempt to
model here. We have argued above that the quantity ${\cal Q}$ is a
measure of the local coherence of the dynamo--generated field.
Thus we approximate these effects by assuming that $B_z$
is subject to an effective magnetic diffusivity $\eta^\ast$ acting in
the radial direction, where

\begin{equation}
\eta^\ast = \eta \max ( 1, {\cal Q} R/H ),
\end{equation}
and {${\cal Q}$ is defined in equation (\ref{Qdef}).

\section{Results}

Even though we have approximated much of the essential physics of a
full MHD disc calculation for reasons of simplicity, we nevertheless
require a number of parameters to specify our results. These are
$\alpha, H/R, B_s, f_{\rm LPP}, k_d = \tau_d(R)\Omega(R)$, and $R_{\rm
out}/R_{\rm in}$. The resulting behaviour of what is a highly
nonlinear system is sufficiently complicated that we do not attempt,
in this paper, to undertake a full survey of parameter space. Rather
we content ourselves with fixing most of the parameters. We take
$R_{\rm out}/R_{\rm in} = 20,000$ and note that the actual value of
this is unlikely to play a significant role provided that it is $\gg
1$. For the runs we discuss below (unless indicated otherwise) we take
$B_s = 10$, since this value seems to ensure that typical values of
the mean poloidal field, $\langle B_z \rangle$, are large enough to
have a noticeable effect on the angular momentum loss, but small enough
(factors of a few times less than the disc field $B_{\rm disc}$) to be
physically plausible. Similarly, unless stated otherwise, we take
$f_{\rm LPP} = 3$, since this value ensures that the physical effect
of the jet torque plays some role without being unphysically dominant. 
We also choose parameters such that $H/R \simeq 0.08$ (see below). We 
then experiment with varying the remaining parameters $\alpha$ and 
$k_d = \tau_d(R)\Omega(R)$.

\begin{figure*}
\includegraphics[width=176mm]{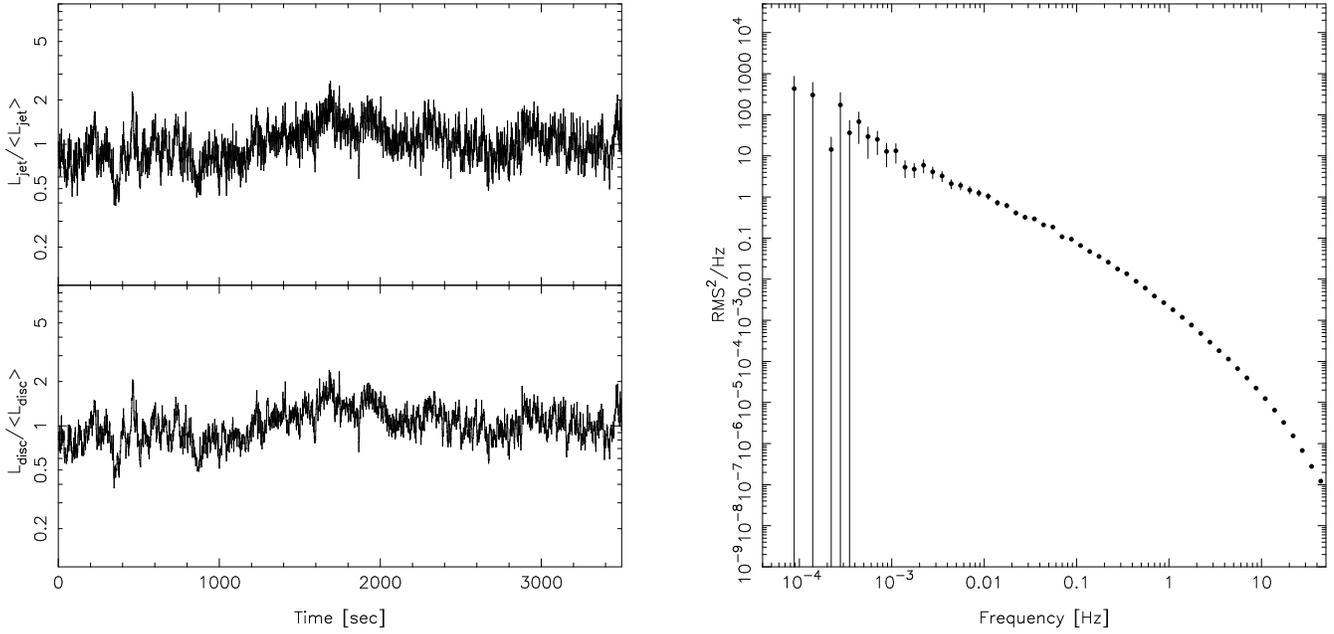}
\caption{The left panel (a) shows representative jet (top) and disc
(bottom) lightcurves for $k_d=10$, $\alpha=0.06$. The lightcurves are
normalised by the mean luminosity in the segment shown, and binned to
10s time resolution. The right-hand panel (b) shows the disc power
spectrum calculated from a 16000s lightcurve. In this and all
subsequent figures we give $1\sigma$ error bars. $<L_{\rm
jet}>/<L_{\rm disc}> = 1.37$ for this case.}
\end{figure*}

In each run, we start from an initial condition in which the disc
surface density is constant at all radii and equal to the arbitrary
value of unity ($\Sigma = 1)$. [Note that we can do this because $\nu$
is independent of surface density, hence the evolution equation
(\ref{eqsigma}) for $\Sigma$ is linear in the absence of wind/jet
angular momentum.]  We start with vertical magnetic field $B_z = 0 $
everywhere, but then take one initial step in the Markoff process at
each annulus.

In order to get some idea of observable disc parameters we compute
estimates of what we denote as a `disc luminosity' $L_d$ and a `wind/jet
luminosity' $L_{\rm jet}$, as functions of time. We regard the disc
luminosity as some measure of the  
radiative emission from the accretion flow. However we do not model
the conversion of gravitational energy into radiation, and the
consequent division of this energy between hard and soft X--ray
components.
It may be that the instantaneous wind/jet luminosity is generally
unobserved, except perhaps when rapid variations give rise to internal
shocks in the outflow (see, however, Malzac 2003).

We take the disc luminosity as caused solely by viscous dissipation
in the disc, and hence (e.g.\ Pringle, 1981) take each disc
annulus $D\!R$ to contribute an amount

\begin{equation}
\delta L_d = \frac{9}{4} \nu \Sigma \pi \frac{GM}{R^2} D\!R.
\end{equation}

We estimate the rate at which energy is removed from each annulus by
the wind/jet from the disc by assuming

\begin{equation}
\delta L_{\rm jet} = (-U_R) \pi \Sigma \frac{GM}{R} D\!R.
\end{equation}

Note that these units are arbitrary in the sense that the units for
$\Sigma$ are arbitrary, but that the relative value of the two
quantities does have meaning. The total disc and jet/wind luminosities
are then obtained by summing these over all the disc annuli, although
evidently most of the contribution comes from the innermost radii. In
the figures below we present values for these quantities averaged over
$10^4$ code time units ($\approx 10$~s for a central black hole of 14
$M_\odot$). We note that for our canonical parameter values of $B_s =
10$ and $f_{\rm LPP} = 3$, the disc and jet luminosities are approximately
scaled versions of each other.

We emphasize that we have implicitly ignored the magnetic energy
represented by $B_z$, which is created/destroyed by the dynamo
process, and is dissipated by the diffusivity $\eta$. One reason for
doing this is that we have no proper model for the processes that
generate and destroy $B_z$ or that determine its physical
scale. However, even when $B_{\rm disc} \sim B_z$, the local energy
flow (per unit disc area) into and out of the poloidal field component
is of order $Q_{\rm mag} \sim H (B_z^2/4\pi)/\tau_d$ and is therefore
less than the rate at which energy is dissipated in the disc by
viscous effects (which depends on $\alpha \propto B^2_{\rm disc}$) by
a factor of $\tau_d \Omega = k_d \gg 1$. Thus, given the
simplifications we have made, neglect of the poloidal energy is
reasonable at this level to a first approximation.

We take the unit timescale for the computations so that the orbital
timescale at the inner edge ($R = 1$) is $2 \pi$. In order to make
comparisons with observations more obvious, in the diagrams and in the
discussion below we have taken parameters such that $M = 14
M_\odot$, and $R_{\rm in} = 6GM/c^2$ so that $R_{\rm in} = 124 (M/14
M_\odot)$~km and the time unit is $1.016 (M/14 M_\odot)$~ms. Runs are
carried out typically for a time of $1.6 \times 10^7 (M/14
M_\odot)$~ms.

\subsection{Timescales and frequencies}

\begin{figure*}
\includegraphics[width=176mm]{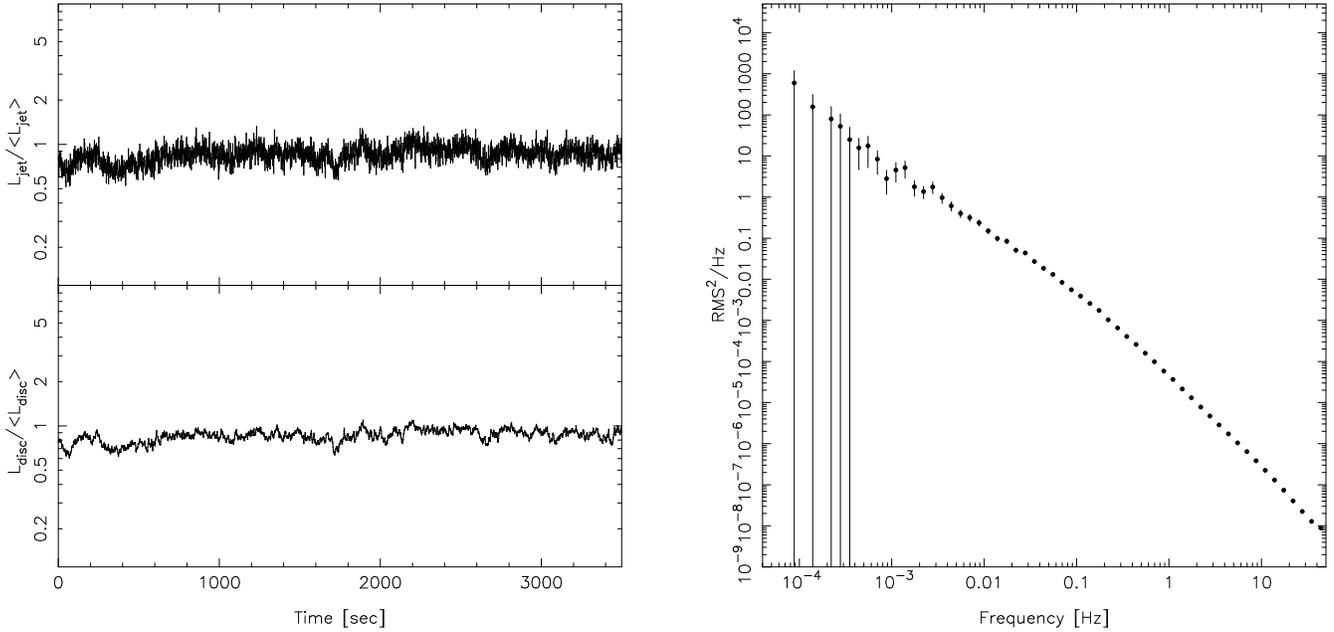}
\caption{Lightcurves (a, left) and power spectrum (b, right) for $k_d=10$,
$\alpha=0.006$. Lightcurves have been normalised and binned as per the
description of Figure~1. $<L_{\rm jet}>/<L_{\rm disc}>=3.38$ for this case.}
\end{figure*}

Before presenting the results of some of our simulations we briefly
discuss some relevant timescales. Thus, we convert code time units to 
milliseconds. For central masses other than 14~$M_{\odot}$ the inverse
angular frequency varies as $\Omega^{-1} \propto M^{-1/2} R^{3/2}$,
thus from $10^{-3} (M/14 \msun)$~s at the inside, $R = 1$, to
approximately $2.8 \times 10^3 (M/14 M_\odot)$~s at $R = 20,000$.
Similarly, the dynamo timescale $\tau_d$, taken equal to $k_d
\Omega^{-1}$, varies as $\tau_d \propto M^{-1/2} R^{3/2}$ from a value
of $0.01 (k_d/10) (M/14 M_\odot)$~s at $R=1$ to a value of $2.8
\times 10^4 (k_d/10) (M/14 M_\odot)$~s at $R = 20,000$. We define the
dynamo frequency as $f_d = \tau_d^{-1}$. The maximum value of this
frequency occurs at the inner edge and is equal to $f_d^{\rm max} =
100 (k_d/10)^{-1} (M/14 M_\odot)^{-1}$~Hz.

The viscous timescale $\tau_\nu$ is given (e.g.\ Pringle 1981) by
\begin{equation}
\tau_\nu \approx \Omega^{-1} (R/H)^2 \alpha^{-1}.
\label{tvisc}
\end{equation}
Thus for the value of $H/R \approx 0.08$ (independent of radius)
adopted in these calculations, the viscous timescale varies as
$\tau_\nu \propto M^{-1/2} R^{3/2} \alpha^{-1}$ from $16(M/14
M_\odot)/(\alpha/0.01)$~s at $R = 1$, to $4.5 \times 10^7(M/14
M_\odot)/(\alpha/0.01)$~s at $R = 20,000$. For the
calculations presented here this means that at times $t \ga 1,000$~s, the
central regions of the disc (say, $R \la 20$) where almost all
the energy is generated, and where all the ``action'' is
taking place, act as if they were being fed at a steady rate by a disc
with unit surface density. The timescale on which magnetic diffusion
works is at most the same as the viscous one, and can be less by a
factor of up to $R/H \approx 12.6$. With the viscous timescale we can
associate a viscous frequency $f_\nu = \tau_\nu^{-1}$. The highest
value of the viscous frequency occurs at the inner edge and for the
disc parameters we use here is given by $f_\nu^{\rm max} = 0.0625
(\alpha/0.01) (M/14 M_\odot)^{-1}$~Hz.

The final relevant timescale $\tau_{\rm mag}$ is that for a 
sufficient number of neighbouring annuli to generate an aligned magnetic 
field, that significant local energy loss to the jet/wind occurs. This 
timescale is discussed in Livio, Pringle \& King (2003). From our
assumptions, the number of zones needed at radius $R$ corresponds to
the number of neighbouring annuli, width $\sim H$, required to span a
radial scale of order $R$. Let this number
be $\bar{N} +1$. Then, assuming for simplicity that $B_z$ can take
only the values $\pm 1$, if each zone acts independently, the
probability of all $\bar{N} +1$ zones being aligned is
$1/2^{\bar{N}}$. To generate a timescale for this we assume that all
the zones at radius $R$ are randomly realigned every $\tau_d(R)$.

To give specific examples, for most of the simulations we take the
number of grid zones to be $N_{\rm grid} = 130$ which gives $H/R =
0.079 \approx 0.08$ and $\bar{N} = 8$. Thus the timescale on which the
innermost 9 zones become aligned is approximately $\tau_{\rm mag}
\approx 2.56 (k_d/10)$~s. Since $\tau_{\rm mag} \propto R^{3/2}$ the
process occurs more slowly at larger radii.
Thus we expect the process to be dominated by timescales at the
smallest radii, but modulated by longer timescales corresponding to
larger radii. The associated magnetic alignment frequency is given by
$f_{\rm mag} = \tau_{\rm mag}^{-1}$. As for the other frequencies the
highest magnetic frequency occurs at the inner edge and is given by 
$f_{\rm mag}^{\rm max} = 0.39 (k_d/10)^{-1} (M/14 M_\odot)$~Hz.

As noted by Livio, Pringle \& King (2003), the quantity $\tau_{\rm mag}$
depends sensitively on $H/R$. For example, if we consider $H/R = 0.135$
we find $\bar{N} = 4$, and $\tau_{\rm mag}$ is reduced by a factor of
16. However in the numerical method of this paper we use the fact that
$D\!R = H$. Thus, changing $H/R$ would require a different numerical
grid, and also change the viscous timescale (for the same value of
$\alpha$). To investigate the effects of changing the magnetic
timescale, without altering other parameters, we here simply vary the
parameter $k_d$. 
Of course in reality $k_d$ is determined by the physics of the disc.
This procedure is designed to mimic the effect of changing $H/R$ in a
numerically convenient fashion. 
Accordingly, we have carried out runs spanning the
parameter space $0.001 \leq \alpha \leq 0.6$ and $10 \leq k_d \leq
1000$.

\begin{figure*}
\includegraphics[width=176mm]{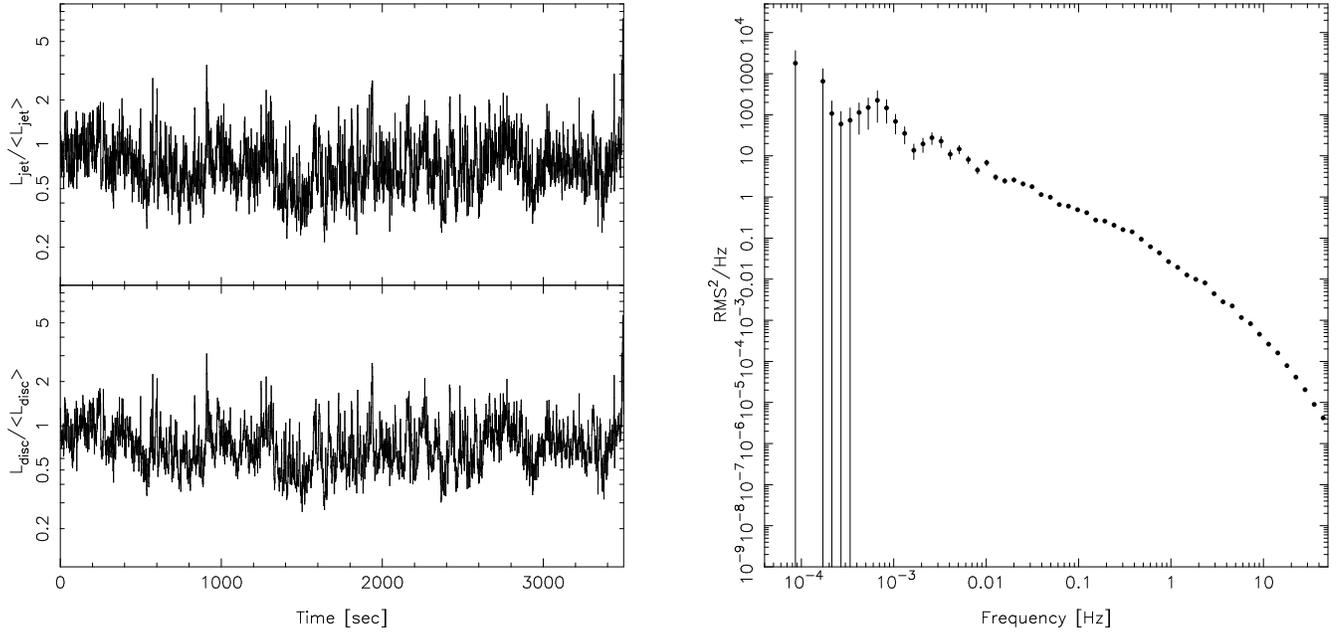}
\caption{Lightcurves (a, left) and power spectrum (b, right) for $k_d=10$,
$\alpha=0.6$. Lightcurves have been normalised and binned as per the
description of Figure~1. $<L_{\rm jet}>/<L_{\rm disc}>=0.50$ for this case.}
\end{figure*}

\subsection{Power spectra}

As we noted above, if $k_d = \tau_d \Omega$ is independent of radius,
and if, as we have assumed, $H/R$ and $\alpha$ are also independent of
radius, then the ratio of the two relevant timescales $\tau_{\rm mag}$
and $\tau_\nu$ is also independent of radius. Thus we expect our
results, appropriately scaled, just to depend on the dimensionless
ratio
\begin{equation}
{\tau_{\rm mag}\over \tau_{\nu}} \sim {2^{R/H}k_d\alpha\over (R/H)^2}
\end{equation}
where we have used (\ref{tvisc}) and $\tau_{\rm mag} \sim \tau_d
2^{R/H}$ (Livio, Pringle \& King 2003).

The power spectrum calculated for each lightcurve is normalised using
the prescription of Miyamoto et~al.\ (1991), namely

\begin{equation}
P_j = \frac{2|S_j|^2}{{\cal R}^2 T}
\end{equation}
where $S_j$ is the Fourier amplitude at frequency $f_j$, ${\cal R}$ is
the mean luminosity of the lightcurve, and $T$ the total duration in
seconds. We have applied a logarithmic frequency binning
to all the PSDs, with $\Delta f/f=0.1$.

\subsubsection{$\tau_{\rm mag} \simeq \tau_\nu$}

In Figure 1(a) we show the light curves for $L_{\rm disc}$ and $L_{\rm
jet}$ and in Figure~1(b) we show the normalised power spectrum of the
disc luminosity for $\alpha = 0.06$ and $k_d = 10$. For these values
of $\alpha$ and $k_d$, $\tau_{\rm mag}(R=1) \simeq \tau_\nu(R=1)
\simeq 2.7/(M/14 M_\odot)$~s, or, equivalently, $f_{\rm mag}^{\rm max}
\simeq f_\nu^{\rm max} \simeq 0.4(M/14 M_\odot)^{-1}$~Hz. From Figure~1(b) 
we see that at frequencies $f \ll f_{\rm mag}^{\rm max} \simeq
f_\nu^{\rm max}$ the power spectrum has the form $p(f) \propto
1/f$. This corresponds exactly to the situation described and
predicted by Lyubarskii (1997). At each radius the local inflow rate
is perturbed randomly with a characteristic timescale which is
approximately equal to the local viscous timescale. These
perturbations then flow inwards on the local viscous timescale, and
when they reach the inner few radii contribute to the total disc
luminosity with a fluctuation on that timescale. The fact that the $1/f$
spectrum extends to frequencies as low as $10^{-3} f_\nu^{\rm max}$
implies that contributions to the variability come from radii as large
as $R \sim 100$. In Lyubarskii's (1997) model, the power spectrum drops
exponentially at frequencies above $f_\nu^{\rm max}$ because his
assumed perturbations have no power at frequencies higher than
this. He reached this conclusion because, as he remarked, he had no
specific physical model for the perturbations. In the model we
consider here there are perturbations to the local inflow rate at all
timescales down to the local dynamo timescale at the inner disc edge,
$\tau_d(R=1)$, corresponding to a frequency of $f_d \simeq 100$ Hz. At
frequencies $f \ga f_\nu^{\rm max}$ the only perturbations affecting
the disc luminosity are those which take place at radii close to the
inner edge where the potential well is deepest.  Thus the shape of the
power spectrum in this frequency regime depends simply on the power
spectrum of the applied random perturbations. From Figure~1(b) it is
apparent that in the model presented here the power spectrum steepens
for frequencies above about $f_{\rm br} \approx 0.2$~Hz $\approx
f_{\rm mag}^{\rm max}$.  The spectrum has a slope of ${\rm d}\log p
/{\rm d} \log f = -2$ at $f \approx 2/(M/14 M_\odot)$~Hz, and $= -3$
at $f \approx 20/(M/14 M_\odot)$~Hz. We see a similar behaviour,
suitably scaled in frequency space, for the run with $\alpha=0.006$
and $k_d = 100$.

\begin{figure*}
\includegraphics[width=176mm]{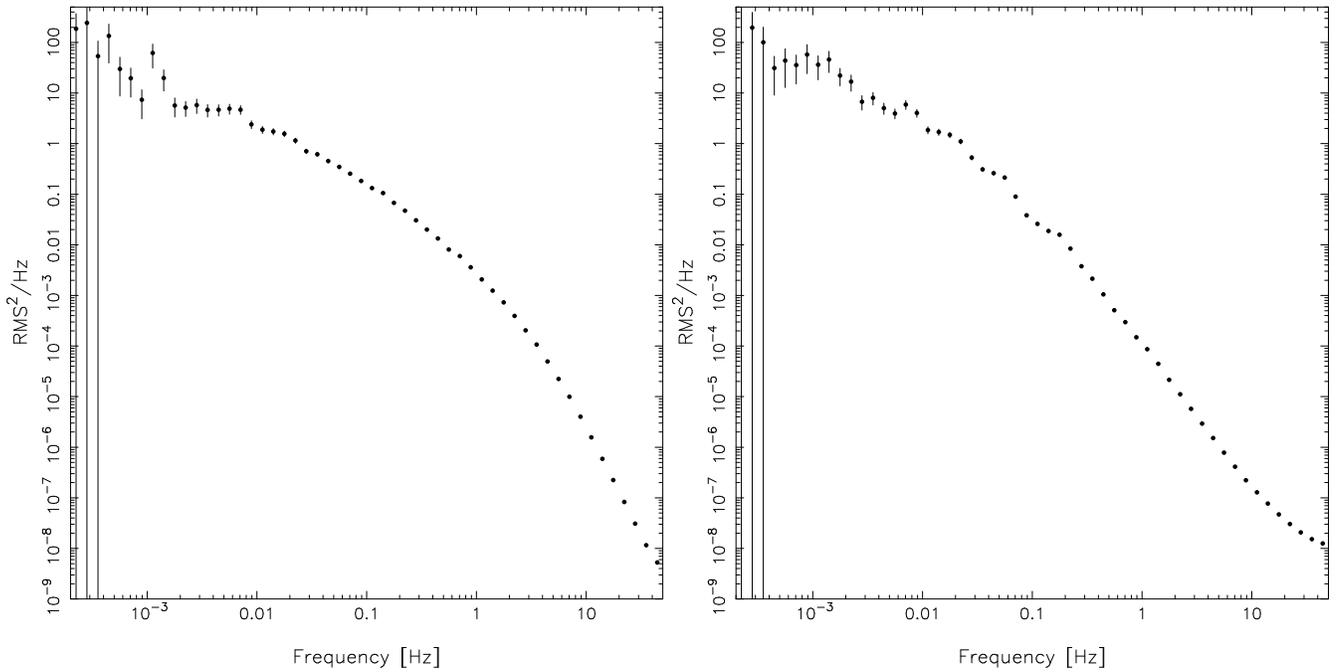}
\caption{Power spectra for $\alpha=0.06$, implying $f_\nu^{\rm max} =
0.4$~Hz, and the cases $k_d=100$ (left) with $f_{\rm mag}^{\rm max} =
0.04$~Hz and, $k_d=1000$(right) with $f_{\rm mag}^{\rm max} = 0.004$}
\end{figure*}

\subsubsection{$\tau_{\rm mag} \ll \tau_\nu$}

We show in Figures 2(a) and 2(b) the light curves and power spectra
for a similar model with $k_d = 10$, but now with $\alpha =
0.006$. Thus in this case we have $f_{\rm mag}^{\rm max} = 0.4/(M/14
M_\odot)$~Hz $\gg f_\nu^{\rm max} = 0.04/(M/14 M_\odot)$~Hz. As
mentioned above, we have the ratio $f_\nu^{\rm max}/f_{\rm mag}^{\rm
max} = 0.1$ independent of radius. In this case the power spectrum is
nowhere of the form $p(f) \propto 1/f$, and indeed is quite well
represented by $p(f) \propto 1/f^2$ at all frequencies. We see similar
behaviour for the case $k_d = 10$, $\alpha = 0.001$, for which
$f_\nu^{\rm max}/f_{\rm mag}^{\rm max} = 0.017$.  In this case the
magnetic torque fluctuations throughout the disc occur locally much
faster than the timescale on which viscous forces can react and inflow
can occur. This implies that the magnetically induced variability has
only a local effect. Thus the disc luminosity, which is dominated by
contributions from the inner regions, is only affected by magnetic
torque fluctuations in those inner regions.

\begin{figure}
\includegraphics[width=85mm]{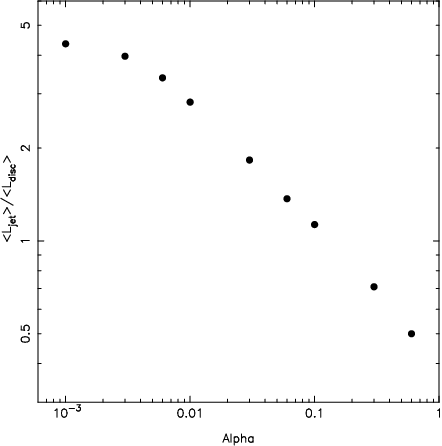}
\caption{The ratio $<L_{\rm jet}> / <L_{\rm disc}>$ versus the
viscosity parameter $\alpha$, for $k_d=10$.}
\end{figure}

\subsubsection{$\tau_{\rm mag} \gg \tau_\nu$}

In contrast with the above, in Figures 3(a) and 3(b) we show the light
curves and power spectra for a similar model, with $k_d =10$, except
that $\alpha = 0.6$. Thus in this case we still have $f_{\rm mag}^{\rm
max} = 0.4/(M/14 M_\odot)$~Hz, but now $ f_\nu^{\rm max} = 4/(M/14
M_\odot)$~Hz $\gg f_{\rm mag}^{\rm max}$. In this regime the power
spectrum is approximately of the form $p(f) \propto 1/f$ for
frequencies $f \la f_{\rm br}$, where $f_{\rm br} \approx 2 f_{\rm
mag}^{\rm max}$, independent of the value of $f_\nu^{\rm max}$. In the
frequency range $f_{\rm br} < f < f_\nu^{\rm max}$, the power spectrum
is roughly of the form $p(f) \propto 1/f^2$, and the power spectrum
falls off more steeply at higher frequencies $f > f_\nu^{\rm
max}$. These findings are also illustrated by the cases $\alpha =
0.06, k_d = 100$ and $\alpha=0.06, k_d = 1000$ (Figure~4). In this
case, because the viscous timescale is short, all magnetic torque
fluctuations can be communicated to small radii, and so can contribute
efficiently to fluctuations in the disc luminosity. At frequencies
lower than about $f_{\rm mag}^{\rm max}$ the situation is still
essentially as described by Lyubarskii (1997) and the power spectrum
is of the form $p(f) \propto 1/f$. However, at frequencies higher than
this, but still below $f_\nu^{\rm max}$, the fluctuations carried
inwards to small radii have ever lower amplitudes, so the power
spectrum steepens. At frequencies higher than $f_\nu^{\rm max}$
however, there is no time for fluctuations to be carried inwards from
larger radii, and all we see is the effects of local fluctuations
occurring in the innermost disc radii. Hence the power spectrum
steepens still further.

\begin{figure*}
\includegraphics[width=120mm]{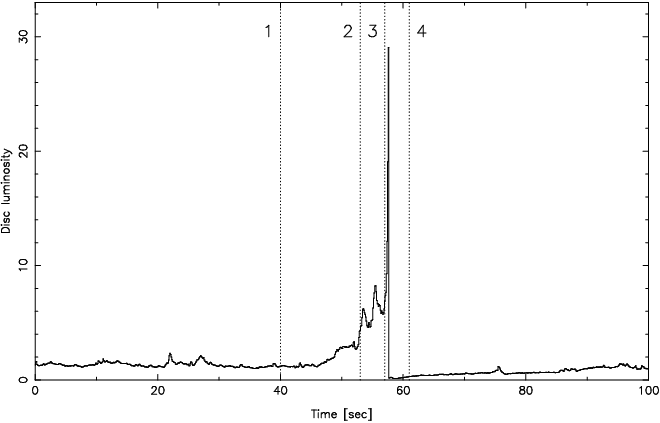}
\includegraphics[height=150mm]{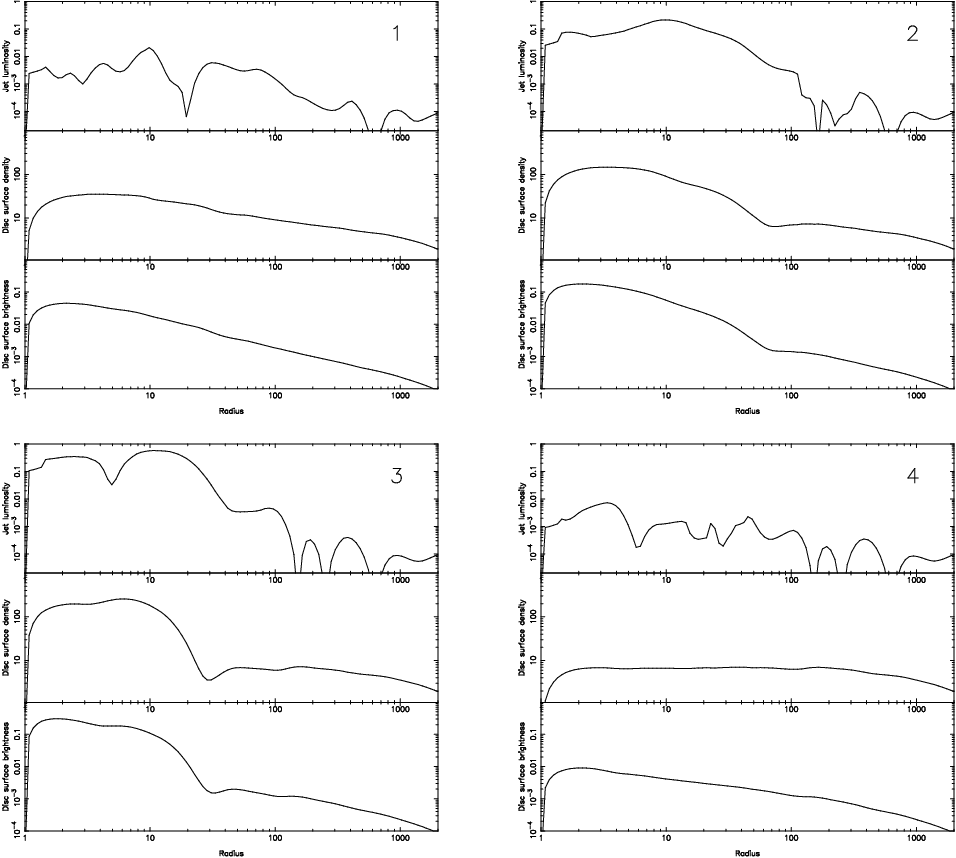}
\caption{The evolution of the disc and jet throughout a large
amplitude flare ($\tau_{\rm d} = 100, \alpha = 0.6$).
Four selected times are marked on the lightcurve. The
top panel shows the disc lightcurve segment covering the flare
event. The four lower panels show the disc surface brightness,
surface density and jet luminosity for the four selected times.}
\end{figure*}


\subsubsection{Effect of the outer disc edge}

In the computations presented so far, all the processes have varied
with radius in a self--similar manner. Thus, for example, the ratio
$f_\nu/f_{\rm mag}$ was independent of radius. For this reason, the
structure of the power spectra has been relatively straightforward to
interpret. In general, of course, the situation is likely to be more
complex, with, for example, both $f_\nu$ and $f_{\rm mag}$ being
intricate, and different, functions of radius. In this Section we
illustrate one simple complication, by setting the dynamo wind
activity to zero outside some radius, $R_{\rm max}$. This radius, can
be viewed for example as the outer edge of the disc. Alternatively, it
can be regarded as a radius outside which the magnetic alignment
timescale, $\tau_{\rm mag}$ becomes very long, because the disc
thickness suddenly drops outside that radius. 

In Figure~\ref{flatpsd} we show the power spectrum for a run which is
identical to that shown in Figure~1 (with $k_d = 10$ and $\alpha =
0.06$), except that dynamo wind activity is turned off for radii $R >
R_{\rm max} = 100$. Since we have assumed that the presence of a
wind/jet depends on magnetic alignment activity over about a factor of
two in radius, we should expect this to give rise to a feature in the
power spectrum at a frequency related to $f_o \approx f_\nu(R_{\rm
max}/2) = f_{\rm mag}(R_{\rm max}/2) = 0.0013$ Hz.  As can be seen by
comparing Figures~1 and ~\ref{flatpsd}, the effect of truncating
wind/jet activity at $R=100$ is to cause a flattening of the power
spectrum from $P(f) \propto 1/f$ to $P(f) \approx$ const. at a
frequency $f \approx 0.002$ Hz.

\begin{figure}
\includegraphics[width=85mm,angle=-90]{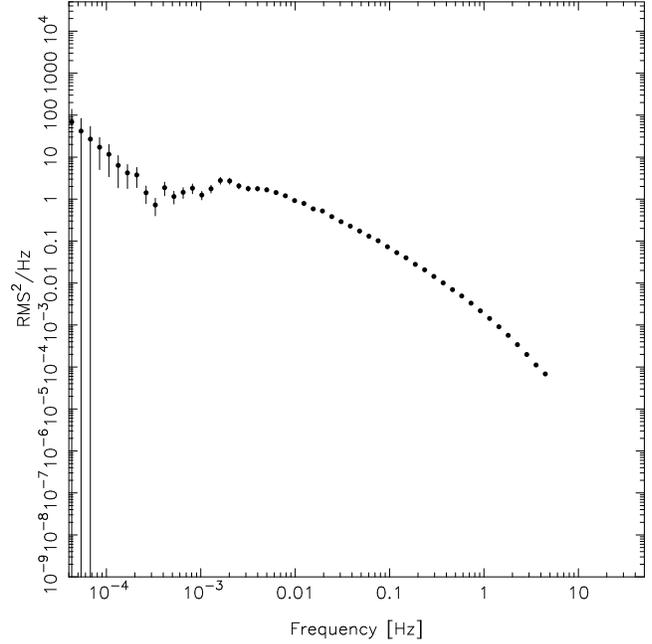}
\caption{The power spectrum for the case $\alpha=0.06, k_d = 10$, for
which the magnetic wind activity has been turned off at radii $R \geq
100$, corresponding to a frequencies of $f < f_o \approx 0.0013$ ~Hz.}
\label{flatpsd}
\end{figure}


\subsection{Lightcurves}

In Figures 1 -- 3, we have shown some typical samples of lightcurves
for both disc and wind/jet emission for $k_d = 10$ and for $\alpha =
0.06, 0.006$ and $0.6$. From the Figures it is evident that the disc and
wind/jet luminosities follow each other fairly closely, except that
the wind/jet emission is much less smooth. This comes about because
the wind/jet emission depends directly on the local magnetic
field configuration, whereas the disc emission depends on the local
disc mass flux which is also mediated by the viscosity. 

\begin{figure*}
\includegraphics[width=176mm]{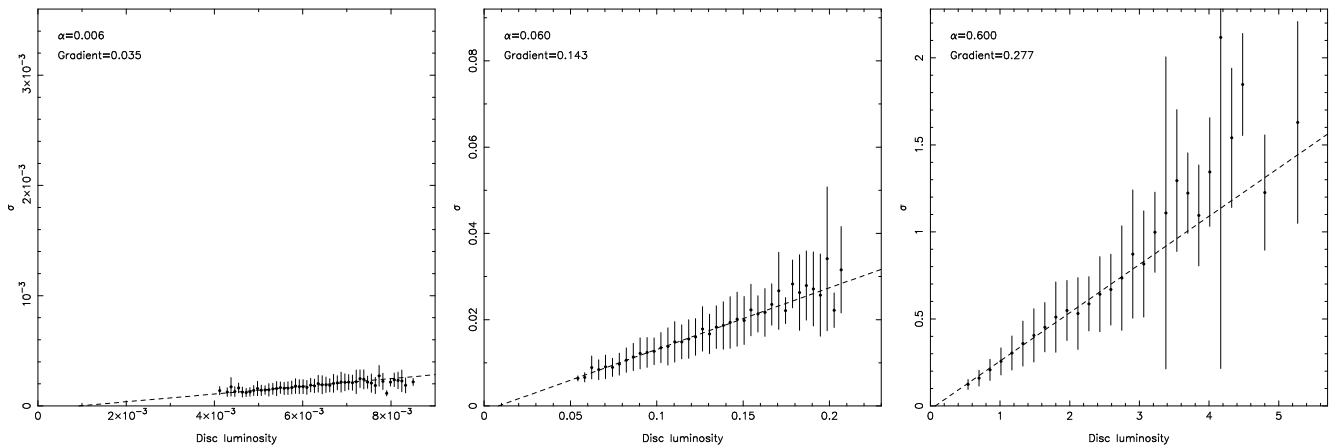}
\caption{Correlation of the rms variability with flux for three
models: $\alpha=0.006$ (left), $\alpha=0.06$ (centre) and $\alpha=0.6$
(right). $k_d=10$ in all cases.}
\end{figure*}

>From the simulations we find, as we increase $\alpha$, that the
amplitude of the flickering increases (see Figures 1 -- 3). We discuss
this in Section~\ref{mchardly} below. We also find that the ratio of
average jet luminosity to average disc luminosity decreases (See
Figure~5). As can be seen from Figure~5, for the parameters we have
chosen $\langle L_{\rm jet} \rangle / \langle L_{\rm disc} \rangle$ is
a slowly varying function of $\alpha$ and is unity for $\alpha \approx
0.1$. Since the value of the poloidal field is scaled to ensure that
the inflow caused by the wind/jet is on average similar to that caused
by the viscosity, it is not immediately obvious what causes this
effect. One possibility is that as $\alpha$ increases, keeping the
dynamo timescale (measured by $k_d$) the same, the wind/jet is able to sweep
material, and magnetic flux, inwards more efficiently, and therefore to do so
at lower values of the poloidal magnetic field.


\subsubsection{Large flares}

In addition to the continual small-scale flickering shown by our
simulated lightcurves, we also see occasional large amplitude
flares. In Figure~6 we show an example of one of these, for a run with
the parameters $\alpha = 0.6, k_d = 100$. Evidently this flare has a
cause similar to that of the small amplitude flickering, except that
it is instigated from a much larger radius ($R \sim 100$), is much
more coherent, and so is of much larger amplitude. Such large flares,
which start at large radius, occur seldom because the timescales at
large radii are correspondingly longer. The Figure shows the disc
luminosity light curve (we note that the jet luminosity is similar but
more spiky), as well as the state of the disc at four particular
epochs (marked in the luminosity plot). In these four plots, we note
that the disc surface brightness and disc surface density are simply
proportional to each other, and are relatively smooth, whereas the jet
surface brightness is much more erratic, reflecting the stochastic
nature of the local poloidal magnetic field. Box~1 and Box~4 show the
disc pre-- and post--flare. Comparison of these two Boxes shows that
the flare depletes the disc at radii $R \la 100$. The timescale for
the recovery of the disc luminosity post-flare indeed corresponds to
the viscous timescale at a radius of $R \sim 100$. We note that
observations of GRS~1915+105 (Belloni et al., 1997) show evidence for
the viscous refilling of a disc depleted by flares. Box~2 shows the
state of the disc at the start of the flare. The incipient dip in
surface density at around $R \sim 80$ is the result of a coherent
magnetic structure, seen in the jet surface brightness, starting an
avalanche of material inwards. Box~3, close to the peak of the flare,
shows that this wave of material has moved further inwards, and that
the surface density in the inner disc has risen accordingly. We note
that this behaviour demonstrates the ability of angular momentum
losses in a disc wind/jet to give rise to a wave--like surge of
inwardly propagating material, in contrast to the diffusive behaviour
of the more usual transfer of angular momentum through viscosity. This
can be seen clearly in Equation~(1) which governs the evolution of
surface density. There, the wind/jet term involving $U_R$ contains a
single radial derivative and so is wave-like, or hyperbolic, in
character, whereas the viscous term, involving $\nu$ has a double
radial derivative and so is diffusive, or parabolic, in character. The
idea that angular momentum loss to a wind might produce such
avalanche/flaring behaviour has been discussed previously by Lubow,
Papaloizou \& Pringle (1994b), by Lovelace, Romanova \& Newman (1994),
and by Livio, Pringle \& King (2003).


\subsubsection{Dependence of flickering amplitude on luminosity}
\label{mchardly}

Uttley \& McHardy (2001) have shown that in some AGN there is a linear
correlation between the amplitude of the flickering and the X--ray
flux. To test whether similar behaviour is evident in our model, we
divide our lightcurves into 100 second segments and calculate the mean
flux ($\bar{F}_j$) and rms flux ($\sigma_j$) for the $j$th segment as
follows:

\begin{equation}
\bar{F}_j=\frac{1}{N}\sum_i F_{ij},
\end{equation}
and
\begin{equation}
\sigma_j=\frac{1}{N}\sqrt{\sum_i (F_{ij}-\bar{F}_j)^2},
\end{equation}
where the index $i$ runs over all time bins in the segment $j$. The
time bins in our lightcurves are 0.1s duration, and thus the
$\sigma_j$ we calculate can be roughly interpreted as the rms
variability in the frequency range 0.01 -- 10~Hz. The normalised rms
variability is then defined as $\bar{\sigma} = \sigma / \bar{F}$. 

In Figure 8 we show plots of $\sigma$ as a function of disc luminosity
for the three models whose lightcurves we show in Figures~1 --
3. These have $k_d = 10$ and have $\alpha = 0.006, 0.06$ and $0.6$.
In each plot the variability amplitude correlates approximately in a
linear fashion with the luminosity. The normalised rms variability is
then measured by the gradient of this relationship and is therefore
roughly a constant for each value of $\alpha$. Thus, we find
that the normalised rms variablity is 3.5 per cent for $\alpha =
0.006$, 14 per cent for $\alpha = 0.06$, and 28 per cent for $\alpha =
0.6$. This increase of normalised rms variability with $\alpha$ is
also evident from the power spectra shown in Figures~1 -- 3. As
was apparent in our discussion of these power spectra, an increase in
$\alpha$ leads to an increase in the efficiency with which
fluctuations at large radius in the disc are swept inwards to small
radii, and thus to an increase in the amplitude of the variability of
the disc luminosity, which is produced mainly at small radii.

\section{Discussion}

Although the disc behaviour we investigate here is relatively
complex, one needs to remember that even within the limits of
the simple model we have introduced, there are a number of physical
processes which we have not addressed, but which are nevertheless
likely to play an important role in any observed source.

One major omission is any consideration of the thermal behaviour of
the disc. This can have a major impact because the magnetic alignment
timescale is a sensitive function of $R/H$, with $\tau_{\rm mag}
\propto \tau_d 2^{R/H}$ (Livio, Pringle \& King 2003). This has a
number of implications. First, since in most discs the ratio $H/R$ is
not a constant, outflow activity is likely to be strong function of
radius --- for example close to the Eddington limit, the inner parts
of the Shakura-Sunyaev disc (1973) have $H \approx$ const. and hence
$H/R \propto 1/R$, which would imply enhanced outflow activity in the
central disc regions. Indeed, we already know that because jets tend
to be launched mainly from the centres of accretion discs, the outflow
behaviour is likely to be strongly non--self similar (See the
discussion by Pringle 1993; Livio 1996; Price, Pringle \& King
2003). Second, the ratio $H/R$ is in general an increasing function of
the accretion rate, or mean disc luminosity. Thus in any given source
it is likely that outflow activity, and hence the nature and magnitude
of the variability, will depend strongly on the brightness of the
source. This effect comes on top of the one presented in
Section~\ref{mchardly}. Indeed, the data presented by Uttley \&
McHardy (2001) show some indication that the relation is steeper than
linear. Third, a local enhancement of inflow rate caused by outflow
activity also leads to enhanced local dissipation within the
disc. This leads in turn to a local increase in disc thickness, a
marked decrease in $\tau_{\rm mag}$ and thus to enhanced outflow
activity. Thus, it seems likely that local outflow activity is
significantly self--enhancing. This phenomenon would lead to much more
pronounced activity than we find in our current models. Fourth, the
thermal hysterisis behaviour which can give rise to outbursting
behaviour on its own (as proposed for example for GRS1915+105 by
Belloni et~al.\ 1997, and Watarai \& Mineshige 2003) could be
considerably affected by enhanced outflow behaviour particularly
during the hot state when the disc thickness is increased (see also
the discussion by Livio, Pringle \& King 2003).

Apart from thermal considerations, there are also likely to be further
complications which will only come to light when more detailed
numerical simulations of the processes involved become possible. One
possible effect here is the relationship between the strength of the
poloidal magnetic field and the dynamo process. The dynamo process is
thought to be driven by the magneto-rotational instability (MRI;
Balbus \& Hawley, 1991). This instability shuts off if the poloidal
field becomes too strong. Suppose then that the dynamo activity at
large radius produces (as we have assumed here) a poloidal field
sufficient to drive an outflow. This causes wave--like inflow in the
disc, carrying with it the poloidal field. As this wave of material
moves to smaller radii, the poloidal field becomes advected with it
and compressed, this process being mitigated by the effective
diffusivity. If at some radius the field becomes strong enough to shut
off the MRI, then at that radius both the effective disc viscosity and
the effective disc diffusivity are significantly reduced. At that
point, the poloidal field is then trapped by the disc, and the
wave--like inflow becomes an unstoppable runaway. If this kind of
behaviour occurs, then the avalanche-like behaviour we have found in
the above models could be a severe underestimate of the violence and
frequency of major flaring behaviour.

\section{Conclusions}

In this paper we have proposed a simple model to describe how the
local dynamo process in an accretion disc, which produces the
effective disc viscosity, can also affect the disc evolution by
driving angular momentum loss in the form of a wind or jet. While the
disc viscosity produces the usual smooth inflow, the accretion rate
driven by the wind/jet is highly stochastic. This comes about because
of our assumption that efficient angular momentum loss to an outflow
can only occur when the local poloidal field produced by the disc
dynamo is sufficiently coherent. We assume that each disc annulus (of
size comparable to the local disc thickness $H$) produces a poloidal
field whose size and direction varies as a random walk with some
characteristic timescale (Markoff process). Only when a number $\sim
R/H$ of such neighbouring annuli produce sufficiently aligned poloidal
fields can outflow and angular momentum loss occur.

We have implemented these ideas in a simple set of equations that
determine the evolution of the disc, and hence the disc and
jet(outflow) luminosities. We have chosen the parameters of the model
($B_s$ and $f_{\rm LPP}$) such that the poloidal field strength
produced by the disc, while smaller than the tangled field within the
body of the disc, is locally large enough to give a significant
outflow. Thus, typically, in our models, $L_{\rm jet}$ and $L_{\rm
disc}$ are comparable in magnitude. From an observational point of
view this is not unreasonable (e.g.\ Fender et~al.\ 2003; D'Elia,
Padovani \& Landt 2003).  We have chosen simple parameters for our
disc models, with constant ratio $H/R$ and constant viscosity
parameter $\alpha$. In addition, we have assumed that local dynamo
timescale $\tau_d$ is a constant $\times$~the local dynamical timescale
$\Omega^{-1}$ in the disc. These assumptions imply that the disc is
self--similar and that the local viscous timescale $\tau_\nu$ is
proportional to the timescale $\tau_{\rm mag}$ on which the local
poloidal field alignment occurs.

We find that in general the disc luminosity shows irregular behaviour.
Small amplitude flickering is superposed on variations in the average
luminosity, together with occasional large amplitude flares. We have
investigated the properties of the flickering with the goal of
obtaining an understanding of what the flickering might be able to
tell us about the properties of the disc. Lyubarskii (1997) showed
that the power spectrum has the form $p(f) \propto 1/f$
if the magnitude of viscosity varies on a timescale everywhere
the same as the local viscous time ($\tau_\nu = 1/f_\nu)$. In our case
the first timescale is that for angular momentum loss in the
outflow $\tau_{\rm mag} = 1/f_{\rm mag}$. This is the case shown in
Figure 1, where $f_\nu = f_{\rm mag}$. The $1/f$ spectrum continues up
to a break frequency $f_{\rm br}$ corresponding to the inner disc
edge, beyond which the spectrum steepens. For
cases in which the local magnetic alignment timescale was longer than
the local viscous time (i.e.\ $f_{\rm mag} \la f_\nu$) we find a
similar behaviour, with the frequency $f_{\rm br}$ corresponding to
the local magnetic alignment timescale at the inner disc edge. We do
not agree with the assumption found in the literature (e.g.\ Vaughan
\& Fabian 2003) that the break frequency corresponds to the viscous
timescale at a particular radius. If the local magnetic alignment
timescale is everywhere shorter than the local viscous timescale, the
spectrum is steeper than $1/f$ throughout (i.e.\ red noise). We have
also shown that the effect of an outer disc edge is to produce a
flattening of the spectrum for frequencies $f < f_o$, where $f_o$ is
the magnetic alignment frequency at that edge.

We have investigated how the amplitude of the flickering varies with
luminosity and with $\alpha$. We find that in any one run, i.e.\ at
fixed $\alpha$, the normalised rms amplitude is approximately
constant, in line with the findings of Uttley \& McHardy (2001) from
observation. We also show that the normalised amplitude of the
flickering increases with increasing $\alpha$, varying in our models
from about 3~per cent for $\alpha = 0.006$ to almost 30 per cent for
$\alpha = 0.6$. This probably comes about because the larger the
viscosity, the more rapidly stochastic behaviour can be swept inwards
from the outer disc regions. Thus, it may be that large amplitude
flickering is a symptom of a large disc viscosity.

We have also demonstrated the possibility of angular momentum loss to
an outflow giving rise to avalanche--like behaviour resulting in a
large flare in luminosity. This is because angular momentum loss
directly to an outflow changes the characteristic nature of the
equation governing the evolution of the disc from diffusive to
wave--like. We note that much of the flickering behaviour also
comes about because of the wave--like nature of the response. This
not only enhances, but also keeps coherent, any perturbations to the
disc.

Finally, we stress again that, although we have applied our findings
to accretion discs around black holes (BHCs and AGN), the whole range
of phenomena found here, and much more (for example, interactions
between the disc and the central object's surface and/or its
magnetosphere), are likely to occur in all systems in which accretion
discs are found.

\section*{Acknowledgments}
ARK gratefully acknowledges a Royal Society Wolfson Research Merit
Award. RGW is supported by the UK Astrophysical Fluids Facility
(UKAFF).  JEP gratefully acknowledges continuing support from the
STScI Visitors' Program.  ML acknowledges support from NASA Grant
GO 7378. We thank Drs.\ P.~Jonker and J.~Malzac for
valuable discussions.

\label{lastpage}
\end{document}